\documentclass{elsart}
\usepackage{graphicx}

\begin{document}

\begin{frontmatter}

\title{Measuring cosmogenic $^9$Li background in a reactor neutrino experiment}

\author[ihep,utsc]{Liangjian Wen},
\author[ihep]{Jun Cao\corauthref{cao}},
  \ead{caoj@mail.ihep.ac.cn}
  \corauth[cao]{Corresponding author. Institute of High Energy Physics,
   Beijing 100049, China. Tel: +86-10-8823-5808; fax: +86-10-8823-3083.}
\author[lbl]{Kam-Biu Luk},
\author[ihep]{Yuqian Ma},
\author[ihep]{Yifang Wang},
\author[ihep]{Changgen Yang}

\address[ihep]{Institute of High Energy Physics, Beijing 100049, China}
\address[utsc]{University of Science and Technology of China, Hefei 230026,
China}
\address[lbl]{Department of Physics, University of California at Berkeley
and \\
Lawrence Berkeley National Laboratory, Berkeley, California 94720,
USA.}

\begin{abstract}
Cosmogenic isotopes $^9$Li and $^8$He produced in the detector are
the most problematic background in the reactor neutrino
experiments designed to determine precisely the neutrino mixing
angle $\theta_{13}$. The average time interval of cosmic-ray muons
in the detector is often on the order of the lifetimes of the
$^9$Li and $^8$He isotopes. We have developed a method for
determining this kind of background from the distribution of time
since last muon for muon rate up to about 20 Hz when the
background-to-signal ratio is small, on the order of a few
percents.
\end{abstract}

\end{frontmatter}

\section{Introduction}
\par
Recently new experiments are proposed to determine the neutrino
mixing angle $\theta_{13}$ using reactor anti-neutrinos with
sensitivity in sin$^{2}2\theta_{13}$ on the order of 0.01
\cite{ihep04}, about a factor of ten better than the current limit
\cite{chooz}. Comparing to the past reactor neutrino experiments
\cite{chooz,paloverde,kamland}, the systematic error of the
proposed experiments will be greatly reduced by carrying a
relative measurement with detectors located at two different
baselines to cancel the reactor-related errors, and other
improvements to the detectors for reducing the detector systematic
errors. However, the cosmogenic $^9$Li and $^8$He background will
remain an irreducible systematic uncertainty in the measurement.

\par
The proposed reactor neutrino experiments detect the
anti-neutrinos via the inverse $\beta$-decay reaction with liquid
scintillator. The inverse $\beta$-decay process is the interaction
of an electron anti-neutrino with a proton (hydrogen) resulting in
the production of a positron and a neutron. A neutrino event is
identified by the coincidence of the energy deposited by the
positron (1 MeV to 8 MeV) followed by the energy released from
neutron capture. Although the proposed experiments will be carried
out underground, a nontrivial number of cosmic-ray muons can still
pass through the hermetically shielded detectors. Some of these
cosmic muons interact with the carbon nuclei in the liquid
scintillator to produce unstable isotopes. Among them, $^9$Li and
$^8$He have $\beta$-neutron cascade decay modes, with branching
fraction of 49.5\% and $(16\pm1)$\%, respectively. The
$\beta$-decay of $^9$Li and $^8$He has a Q value of 13.6 MeV and
10.7 MeV, respectively, which overlaps with the positron signal of
the neutrino events. The beta particle can thus mimic the prompt
signal of a neutrino event while the neutron provides the delayed
signal. Hence the cosmogenic $^9$Li and $^8$He isotopes are
indistinguishable from the neutrino events. Since these two
isotopes are relatively long-lived, with a half-life of 0.178 s
for $^9$Li \cite{li9lifetime} and 0.119 s for $^8$He
\cite{he8lifetime}, they are difficult to reject with a muon veto.

\par
Production of $^8$He and $^9$Li has been measured with 190 GeV
muons on a liquid-scintillator target at CERN \cite{hagner}. Since
their lifetimes are so close it is hard to determine their
individual cross sections. The cross section $\sigma({\rm ^9Li} +
{\rm ^8He})$ is measured to be $(2.12\pm 0.35) \mu{\rm b}$, and
the energy dependence of the cross section is suggested to be
$\sigma_{\rm tot}(E_\mu) \propto E_\mu^\alpha$, with $\alpha =
0.73$. In the KamLAND experiment, about 85\% of the isotopes are
found to be produced by inelastic scattering of energetic muons
\cite{kamland,mckinny}. From the time distribution and the
$\beta$-energy spectrum of the $\beta$-neutron events, the
contribution of $^8$He relative to $^9$Li is less than 15\% at the
90\% confidence level. Furthermore, the $^8$He contribution could
be determined by tagging the cascade $^8$He $\rightarrow ^8$Li
$\rightarrow ^8$Be decay sequence \cite{dchooz}. Hence, for
simplicity, we assume that the cosmogenic isotope background is
solely $^9$Li in our investigation.

\par
While KamLAND can measure the $^9$Li background well from the time
distribution of the $\beta$-neutron event since last muon, it is
not necessarily true for other experiments with less overburden.
The average overburden of the KamLAND detector is 2700 m.w.e.
(meter-water-equivalent), resulting in a muon flux of 0.0015
Hz/m$^2$, or 0.2 Hz of cosmic-ray muons in the active volume of
the detector. This implies the mean time interval of two adjacent
muons is 5 s, much longer than the lifetime of $^9$Li. For most
reactor experiments proposed to measure $\theta_{13}$, the
overburden ranges from about 300 to 1 200 m.w.e. \cite{ihep04}.
The muon rate can be as high as 20 Hz, resulting a mean time
interval of muons significantly shorter than the lifetime of
$^9$Li. However, we find that it is still possible to determine
the $^9$Li content since the anti-neutrino rate and the $^9$Li
rate are significantly lower than the muon rate. The probability
of having two $\beta$-neutron events (either neutrino or $^9$Li
candidates here) in the time window used in the analysis is
negligible. Thus the muons immediately following neutrino
candidates can simply be ignored. In this paper, we show that it
is possible to measure the $^9$Li and $^8$He background in-situ
even when the cosmic-ray muon rate, comparing to the isotope
lifetime, is relatively high.

\section{Method of Least Squares}
\par
Suppose the muons are randomly distributed in time. The time
interval $t$ of two adjacent muons follows a normalized exponential
function
\begin{equation}\label{eqn:dtdis}
f_\mu(t) = \frac{1}{T}\exp(-t/T) \,,
\end{equation}
where the constant $T$ is the mean time interval between two
adjacent muons, $T=1/R_\mu$, with $R_\mu$ being the muon rate. The
neutrino events are also randomly distributed and independent of
the muons. The time interval between a neutrino event and the last
muon preceding it obeys the same exponential distribution, but
with a different constant $T^\prime=1/(R_\mu+R_\nu)$, where
$R_\nu$ is the neutrino rate. The neutrino rate is normally
several orders lower than the muon rate. The difference in
$T^\prime$ and $T$ can thus be ignored. Therefore, we have
$f_\nu(t)=f_\mu(t)$. Here, the variable $t$ is the time of the
neutrino event since the last muon.

\par
The $^9$Li events are correlated with the muons. If the muon rate
is high, a $^9$Li event might be the product of any of the
preceding muons. When summed over the contributions of all
preceding muons, the probability density function (p.d.f.) of the
$^9$Li events in terms of the time since last muon $t$ is
\begin{eqnarray}
f^\prime_{Li}(t) & = & \frac{1}{\tau}e^{-t/\tau}+
 \int\frac{1}{T}e^{-t_1/T}
 \cdot \frac{1}{\tau} e^{-(t_1+t)/\tau} dt_1 + \dots \nonumber \\
 &  & + \int\frac{1}{T^n}e^{-\sum_{i} t_i/T} \cdot \frac{1}{\tau} e^{-(\sum_i t_i+t)/\tau}
  dt_1\cdot\cdot\cdot dt_n + \dots \nonumber\\
& = & \left(\frac{1}{\tau}+\frac{1}{T}\right)e^{-t/\tau} \,,
\end{eqnarray}

where $\tau=0.257$ s is the lifetime of $^9$Li,  and $t_i$ is the
time interval of two adjacent preceding muons. The first term
corresponds to the contribution from the last muon. The second
term is that from the second-to-the-last muon, and so on. Because
the time since last muon is $t$, the time interval between the two
muons immediately before and after a $^9$Li event must be greater
than $t$. This condition has a probability of $\exp(-t/T)$.
Therefore the p.d.f. of the time since last muon $t$ for the
$^9$Li events becomes
\begin{equation}
f_{Li}(t) = \frac{1}{\lambda}\exp(-t/\lambda), ~~~~~~~~~
\frac{1}{\lambda}=\frac{1}{\tau}+\frac{1}{T}\,.
\end{equation}
In a reactor neutrino experiment, the observed distribution of the time
since last muon $t$ for all $\beta$-neutron events is a combination of
the distribution of the genuine neutrino events and that of the $^9$Li
events. As a result, the observed distribution can be described by
\begin{equation}\label{eqn:chi2}
f(t) = B\cdot \frac{1}{\lambda}\exp(-t/\lambda) +
S\cdot\frac{1}{T}\exp(-t/T) \,,
\end{equation}
where $B$ and $S$ are number of $^9$Li and neutrino events,
respectively. The background-to-signal ratio is simply $B/S$. The
amount of the $^9$Li background can then be measured by minimizing
the $\chi^2$ function which is a sum of the squares of the
residuals of the observed time distribution divided in bins from
the expectation given by Eq.~\ref{eqn:chi2}.

\section{Method of Maximum Likelihood}
\par
In the least-squares method, contributions of all preceding muons
are included in the p.d.f. and expressed as a function of the time
since last muon. Similarly, an unbinned maximum-likelihood fit can
be performed with the same p.d.f.. The log-likelihood function
(MLa) is thus defined as
\begin{equation}\label{eqn:mla}
\log{L}
  =  \sum_i\log\left[
 b\frac{1}{\lambda} e^{-t_i/\lambda} + (1-b)\frac{1}{T}
 e^{-t_i/T} \right] \,,
\end{equation}
where $t_i$ is the time of the $i$-th $\beta$-neutron event since
last muon. Here we assume that each $\beta$-neutron event has a
probability of $b$ being a $^9$Li event and $1-b$ to be a neutrino
candidate. To precisely determine $\theta_{13}$, it is desirable
to have the background-to-signal ratio as small as possible. In
this case, $b$ can be taken as such a ratio.

Alternatively, the maximum-likelihood method can exploit the
contributions to a $\beta$-neutron event from each preceding
muons, instead of averaging them into the time distribution since
last muon. When the muon rate is not too high but contributions
from the preceding muons other than the last one cannot be
ignored,
we can construct a log-likelihood function (ML) as
\begin{equation}\label{eqn:LLF}
\log{L}=\sum_i\log\left[b\left(\sum_j
\frac{1}{\tau}e^{-t_{ij}/\tau}\right)e^{-t_{i1}/T}
+(1-b)\frac{1}{T}e^{-t_{i1}/T}\right]    \,,
\end{equation}
where $i$ sums over all $\beta$-neutron events in the sample and
$j$ sums over all preceding muons of the $i$-th $\beta$-neutron
event. Here $t_{i1}$ is the time of the $i$-th $\beta$-neutron
event since last muon, and $t_{ij}$ is its time since the $j$-th
preceding muon. As in the least-squares method, the term
$\exp(-t_{i1}/T)$ is included in the probability density function
of the $^9$Li to account for the condition that there is no muon
between the $\beta$-neutron event and the last muon. In practice,
only muons in a finite time window are included. This truncation
will lead to an error in the probability on the order of $10^{-3}$
for a two-second window.

\par
The statistical error of the parameter in the fit can be estimated
for the maximum-likelihood method when the sample size is very
large. In this case, the variance $V$ of the parameter $\theta$ is
given by~\cite{msda}
\begin{equation}
V^{-1}(\hat{\theta})=\left[ E\left( -\frac{\partial^2\log
f(\vec{\bf X} | \theta)} {\partial\theta^2} \right)
\right]_{\theta=\hat{\theta}}    \,,
\end{equation}
where $f$ is the density function, $\hat\theta$ is the expectation
of the parameter $\theta$, $\vec{\bf X}$ is the sample space, and
$E$ denotes the mathematical expectation in the sample space. For
the MLa given in Eq.~\ref{eqn:mla}, the variance is
\begin{equation}\label{eqn:var}
V^{-1}(\hat{b})
  =  N \int \frac{ \left( \frac{1}{\lambda} e^{-t/\lambda} - \frac{1}{T}
 e^{-t/T}\right)^2}
 {\hat{b}\frac{1}{\lambda} e^{-t/\lambda} + (1-\hat{b})\frac{1}{T}
 e^{-t/T}} \; dt \,.
\end{equation}
If $b \ll \tau/(\tau+T)$, which is the case for most reactor
neutrino experiments, the above integral can be simplified. The
statistical error of $b$ is approximated to be
\begin{equation}\label{eqn:sigb}
\sigma_{b}=\frac{1}{\sqrt{N}}\cdot \sqrt{(1+\tau R_{\mu})^2-1} \,,
\end{equation}

where $N$ is the total event number of $\beta$-neutron events. It
is interesting to note that the error is independent of $b$. When
the muon rate is high, as expected, the error gets worse.

\section{Monte Carlo Simulation}
\par
To validate the formulation, we have simulated data samples of a
reactor neutrino experiment, with neutrino, muon, and $^9$Li
events taken into account. For each sample, two independent data
sets are first generated, one contains neutrino events, and the
other contains muons and $^9$Li events. The neutrino events are
generated randomly in time. For a given muon rate $R_\mu$, the
number of muons is determined. Muons are then generated randomly
in time. The probability of producing the $^9$Li isotope by each
muon is calculated with a Poisson distribution based on the
production cross section measured at 190 GeV. If a $^9$Li is
created, its decay time is generated according to an exponential
function using the $^9$Li lifetime. Neither the energies of the
$^9$Li and neutrinos nor the muon-energy dependence of the yield
of $^9$Li is taken into account here. The two independent data
sets are then combined and sorted by time to form the
$\beta$-neutron sample for analysis.

\begin{figure}[!htb]
\begin{center}
 \includegraphics[width=8cm]{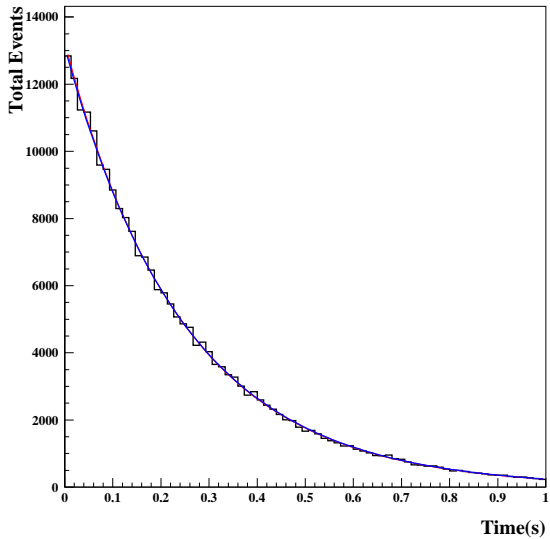}
\caption{Distribution of time since last muon for Monte Carlo
simulated $\beta$-neutron events (a combination of genuine
neutrino and $^9$Li events). The input muon rate is 4 Hz and the
B/S ratio is set at 0.01. The red line is the best fit of Eq.
\ref{eqn:chi2} to the distribution. The blue line is a fit with B
= 0 and S as a fit parameter.} \label{fig:beta-n}
\end{center}
\end{figure}

The distribution of the time since last muon for all the
$\beta$-neutron events in a Monte Carlo sample with an input muon
rate of 4 Hz and a value of 0.01 for B/S is shown in Fig.
\ref{fig:beta-n}. The red curve is the best fit of Eq.
\ref{eqn:chi2} to the distribution, whereas the blue curve is a
fit with B fixed to 0 and S as a fit parameter. Since the amount
of $^9$Li background is small, the two curves are almost
identical.

\par
\begin{figure}[!htb]
\begin{center}
 \includegraphics[width=8cm]{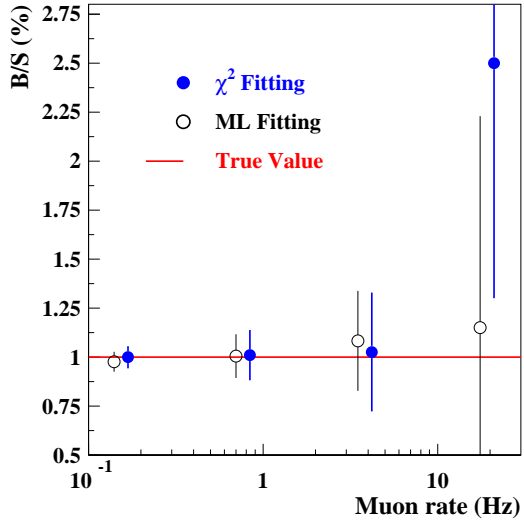}
\caption{Background-to-Signal ratios (B/S) obtained with the
least-squares and maximum-likelihood methods as a function of the
muon rate. The input background-signal ratio is fixed at 1\%. For
each muon rate, $2.5\times10^5$ neutrino events are generated,
corresponding to a 0.2\% statistical error. The results from the
least-squares fit are plotted to the right of those obtained with
the maximum-likelihood fit for clarity.} \label{fig:comp}
\end{center}
\end{figure}

The results obtained with the least-squares and maximum-likelihood
(ML) methods are shown in Fig.~\ref{fig:comp}. For each given muon
rate, the background-to-signal ratio, B/S, is fixed at 1\% since
the statistical error is independent of it (see Eq.
\ref{eqn:sigb}). In addition, 400 samples are generated and
analyzed for each given muon rate to investigate the potential
bias and precision of the fit. It is clear from
Fig.~\ref{fig:comp} that both the least-squares and
maximum-likelihood methods can determine the background-to-signal
ratio reliably up to a muon rate of about 20 Hz.

\par
\begin{figure}[!htb]
\begin{center}
 \includegraphics[width=8cm]{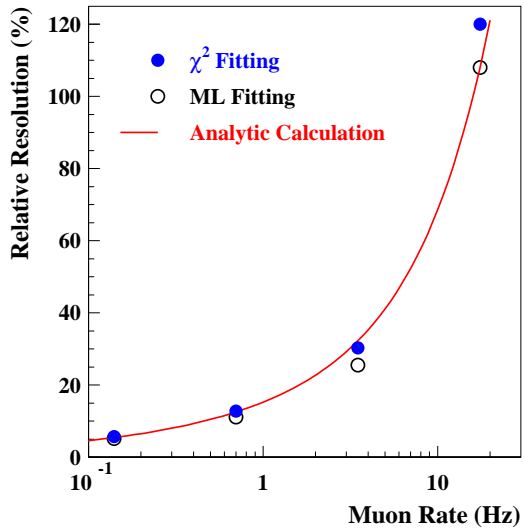}
\caption{Relative resolution of B/S as a function of the muon
rate, The solid line is the analytical result of
Eq.~\ref{eqn:sigb}. The relative resolution is defined as the
statistical error (standard deviation) obtained from the fit
normalized to the input value of 1\% for the B/S ratio.}
 \label{fig:reso}
\end{center}
\end{figure}

Along with the analytical estimation of Eq.~\ref{eqn:sigb}, the
relative resolution of B/S obtained with the least-squares and
maximum-likelihood fits as a function of the muon rate are shown
in Fig.~\ref{fig:reso}. Here the relative resolution is defined as
the statistical error (standard deviation) obtained from the fit
normalized to the input value of 1\% for the B/S ratio. The
results from the two fitting methods are in excellent agreement
with the analytical calculation.

\section{Conclusions}
\par
We have shown that, as long as the occurrence of (anti-)neutrino
events and $^9$Li decays is low comparing with the $^9$Li
lifetime, it is possible to measure in-situ the cosmogenic $^9$Li
background for a reactor neutrino experiment with relatively high
muon rate. The muons immediately following a $\beta$-neutron event
can be safely ignored because they have very small probability to
be associated with another $\beta$-neutron event. We have obtained
probability density functions that can be utilized for determining
the amount of $^9$Li background using the least-squares or
maximum-likelihood fit, with the latter method having a slightly
better precision. When the background-to-signal ratio $b$ is small
such that $b \ll \tau/(\tau+T)$, the precision of the in-situ
measurement can be estimated with a simple formula in Eq.
\ref{eqn:sigb}, which depends on the muon rate and the number of
$\beta$-neutron events in the sample.

\section{Acknowledgements}
Y.F.W. was supported by grants from the National Science Funds for
Distinguished Young Scholars (10225524). K.B.L. was grateful for
the hospitality of IHEP during sabbatical visit, and was partially
supported by the US Department of Energy's Office of Science.


\begin{thebibliography}{00}
\bibitem{ihep04}
For example, see Y. Wang,
Recent results of non-accelerator-based neutrino experiments,
Proceedings of ICHEP 2004, vol. 1, 188. [hep-ex/0411028]. K. Luk,
Determining $\theta_{13}$ using nuclear reactors, ibid, 204.

\bibitem{chooz}
M. Apollonio et al., Eur. Phys. J. {\bf C27} (2003), 331.

\bibitem{paloverde}
F. Boehm et al., Phys. Rev. {\bf D62} (2000), 072002.

\bibitem{kamland}
K. Eguchi et al., Phys. Rev. Lett. {\bf 90} (2003), 021802; T.
Araki et al., ibid, {\bf 94} (2005), 081801.

\bibitem{li9lifetime}
Booth et al., Nucl. Phys. {\bf A119} (1968) 233; Abramov et al.,
Sov. J. Nucl. Phys. {\bf 23} (1976) 636.

\bibitem{he8lifetime}
F. Ajenberg-Selove, Nucl. Phys. {\bf A490} (1988) 1.

\bibitem{hagner}
T. Hagner et al., Astro. Phys. {\bf 14} (2000), 33.

\bibitem{mckinny}
K. S. McKinny, Ph.D. thesis (unpublished), University of Alabama, 2003.

\bibitem{dchooz}
F. Ardellier et al., hep-ex/0405032.

\bibitem{msda}
For example, see John A. Rice,
Mathematical Statistics and Data Analysis (second edition),
Wadsworth Publishing Co. Inc. (1993).


\end{thebibliography}
\end{document}